# Educational asymmetries in the making: Science Fair Competitions as Proxies of the Economic Development

*Dejan Vinković*

Physics Department, Faculty of Natural Sciences and Mathematics, University of Split
vinkovic@pmfst.hr

*Dunja Potočnik*

Institute for Social Research in Zagreb
dunja@idi.hr

ABSTRACT   Croatia does not score well in the human development index, rate of employment, or development of the high tech sector – to mention only a few indicators that hinder good prospects for improvement. Also, Croatian high school students are at the bottom of European rankings on performance in science and mathematics and their interest in career in science is declining. Croatia needs more educated people, especially in the area of science and technology, and this can be achieved only if the youth become interested in science at an early age. Thus, science fair competitions are of an utmost importance for development of Croatia. This paper aims at investigating a relation of the Croatian counties' development index and their students' participation rates in the science fair competitions. This is done by including two development indexes in search for the relation with the counties' participation rates in biology, chemistry, physics and mathematics competition, and with a reference to a previous research on astronomy competition. As is revealed, there is a growing trend of interconnection of the development index and participation rates at science competitions in all disciplines.

## 1. Education and Modern Economies – Croatian Perspective

The increase of proportion of the highly educated youth and their employability can be seen as one of the pillars of reaching societal and economic stability in Croatia (*Science and Technology Policy of the Republic of Croatia 2006-2010*, 2006). It is not only crucial to have educated youth – more sound orientation towards science and technology (S&T) is needed. Although Croatia is not the only one lacking highly educated people in S&T, Croatian position is critical because of its low score on several indicators of educational, economic and technological advancement. In their research on young people's interest in S&T, Sjøberg and Schreiner (2006) refer to the *Europe needs more scientists* (2004) document, which gives a special attention to the number of people entering S&T education and careers in the European Union. The report title reveals the key issue: the falling recruitment of students to S&T is seen as a major problem in most European countries. As shown by the Eurobarometer survey *Europeans, Science & Technology* (2005) youth interest in S&T is on a constant decline. Recent research on career aspirations of teenagers (Schoon and Parsons, 2002) is conceptualized in the context of the theories that recognize the impact of contextual factors (Lent, Brown and Hackett, 2002), such as the exposure of teenagers to different socialization practices. These factors include psychological, historical, cultural, economic and socio-political variables.

Sjøberg and Schreiner are authors of one of the ROSE[1] publications that takes the young people's values, views and ways of understanding themselves, their surroundings and the world in which they are growing up. Accordingly, their data show a strong relationship between the human development index[2] (HDI) of a country and the response in the ROSE questionnaire. For example, high school pupils in the Scandinavian countries and Japan (countries with the highest HDI) express the lowest interest for careers in science and technology among all the countries included in the study. On the other hand, almost all students in the least developed countries, like Bangladesh, Uganda, Zimbabwe, etc., want to become scientists. One questionnaire item asked the students whether they agreed that school science is interesting. Compared to the less economically-developed countries, the students in the more economically-developed countries showed little interest in the subject. Moreover, neither girls nor boys in Western societies particularly want to be scientists.

---

[1] The Relevance of Science Education is an international comparative project designed to shed light on affective factors of importance to the learning of science and technology. The target population is comprised of students towards the end of secondary school (age 15).

[2] The Human Development Index (HDI) is an index used to rank countries by level of "human development", which usually also implies whether a country is developed, developing, or underdeveloped. The HDI has been used since 1990 by the United Nations Development Programme and it combines three dimensions: I) life expectancy at birth, as an index of population health and longevity; II) knowledge and education, as measured by the adult literacy rate (with two-thirds weighting) and the combined primary, secondary, and tertiary gross enrollment ratio (with one-third weighting) and III) standard of living, as measured by the natural logarithm of gross domestic product per capita at purchasing power parity.

Young people wish to be passionate about what they are doing and they wish to develop themselves and their abilities. If they have a range of possible and accessible futures they will, among the many options, choose the most interesting. Publication *Science Education Now* (2007) by European Commission recognizes a need for a change in teaching methods in order to make S&T more appealing to young people. It proposes a reversal of school science-teaching pedagogy from mainly deductive to inquiry-based science education that has proved its efficacy in increasing children's and students' interest in S&T. At the same time this method also stimulates teachers' motivation.

Croatia has recognized some of the above mentioned educational goals as the basic prerequisites for its desire to develop a knowledge-based economy (*Strateški okvir za razvoj 2006.-2013; Znanstvena i tehnologijska politika Republike Hrvatske 2006.-2010., 2006*). On the other hand, according to the 2001 Census, it has only 15,8% of tertiary educated people and only 0,23% of doctorate holders (age group of 25-64). The population with tertiary education attainment in Croatia is concentrated in major towns, especially in Zagreb – Zagreb has 18% of the Croatian population, but accounts for 34% of the population with tertiary educational attainment and 56% of scientific master and doctoral degree holders in Croatia. This has long term consequences on the development of Croatia, since Zagreb creates asymmetries in the educational system. For example, it attracts high school students from the nearby geographical regions and, therefore, has around 6.500 more high school than primary school students (Central Statistical Bureau, 2008). Lučin (2007) reveals that in order to achieve results comparable to those of more advanced European economies by the year 2025, Croatia would need to have at least 40% of people in the age group of 25-64 with first or second degree of higher education, whereas a number of the doctorate holders should increase to at least 2% (i.e. tenfold). Croatia also needs a turn toward more knowledge-intensive services, which requires an advancement of natural and technical sciences (*Science and Technology Policy of the Republic of Croatia 2006-2010*, 2006). Contemporary trends do not go along these recommendations. For example, Babić *et al*. (2006.: 146) show the number of university students per science areas and find that there has been an increase of 143% in social sciences and humanities since 1991, whereas biotechnology has recorded a decrease of 24%, while other areas of S&T have an increase between 20% and 38%. As shown by the research on the Zagreb University students (Potočnik, 2009), 89% of students enrolled into a given university programme out of personal interest into the subject. Hence, if we want an increase in S&T enrollment then serious steps are required toward rising the interest of youth in S&T. We can also add that according to a Eurostat study (Meri, 2008.:51) Croatia is lagging behind Europe in the number of people employed in high-tech sectors. The study shows that Cyprus and Estonia are at the top with almost 60% of people with tertiary education in high-tech sectors. Belgium, French, Norway with almost 50% follow, while Finland, Sweden, Denmark, Bulgaria, Lithuania and UK have around 40%. European average is 36% while Croatia has 28%, sharing its position with Turkey, Czech Republic and Macedonia.

Western European standard of living within the next two decades is sustainable if the societal consensus or sound macroeconomic management and continuing orientation towards rapidly developing human capital are maintained (Ederer; Schuler; Willms, 2007). For the countries that scored poorly on the European Human Capital Index (Ederer; Schuler; Willms, 2007) – namely, Bulgaria, Croatia and Poland – "there is a realistic chance of being stuck in relative poverty to the European average – since no other resource but human capital can lift them out of the situation they are in today". Among the named countries the utilisation of human capital is exceptionally low; on average it falls 7% lower than the EU-14 average. The spread between some of the worst performers in human capital utilisation – such as Croatia, Poland and Slovakia – and Western Europe's best utilizers of human capital – such as Denmark and the Netherlands – amounts to around 17 percentage points. The inability to integrate this invested human capital into the value-creation cycle is a severe waste of economic development potential in these countries.

Additional data on Croatian incapacity to deal with the modern economies can be found in the 2008 Eurostat Yearbook. The data say that employment in high- and medium-high-technology manufacturing is: EU average 5,6% (lowest Latvia 2,5%, highest Germany 9%), Croatia 4,4%; while employment in knowledge-intensive services is: EU average 32,6% (lowest Romania 14,5%, highest Sweden 47,5%), Croatia 22,1%. Moreover, *The Global Competitiveness Report* puts Croatia on $61^{st}$ place among 134 countries included in the study, rating the higher education and training as $48^{th}$, quality of educational system as $66^{th}$ and quality of math and science education as $30^{th}$. The other data for Croatia in this report are also inadequate for better prospects in education and research and development (R&D): its capacity for innovation is at the $42^{nd}$ place, quality of scientific research institutions – $50^{th}$, company spending on R&D – $45^{th}$, university-industry collaboration – $43^{rd}$, availability of scientists and engineers – $58^{th}$ and utility patents – $35^{th}$ place.

## 2. Science Fair Competitions and Economic Development

In the previous chapter we offered some insights into the inability of the Croatian educational system to deal with a growing need for S&T development. In this part we will further elaborate why it is dangerous to let current trends in education and economy persist. We will also give a rationale behind our survey and a methodology overview. For a start, we consider a research conducted at the Institute for Social Research in 2004 (Baranović, 2006). Its results give insights into the popularity of some courses in the primary school, as well as the most frequent teaching and learning methods in these courses. According to this research natural science courses are at the bottom of popularity; biology is a favourite natural science course in the primary school whereas physics and chemistry are at the bottom. The reasons for this (un)popularity as stated by the pupils are (dis)interest, where a role of teachers and utility for the future is mentioned in the case of physics, while course content and activities are the stated reasons for interest in biology. We can say that the teaching methods play a crucial role in creation of attraction of students toward natural science courses. But, it seems that the teachers are not aware enough of this – 73% of the pupils said they never or rarely presented their work in a shape of seminar or other type of independent work; 91% never or rarely write comments, reports or observations; 80% never or rarely took part in a field work; 68% never or rarely participated in making of some artefacts like a poster or a model. On the other side, 56% passively sit and write down the teachers' words. As we saw from the example of ROSE results, interest in the natural and technical sciences declines with the increase of the country's HDI. As we can expect a long term increase of the HDI in Croatia, we can accordingly expect further decline in the interest of young people for professions in natural and technical sciences. In a combination with a fact that the Croatian educational system lacks hundreds of teachers of mathematics and physics and that many of those who are currently working will be retired in a couple of years we can see that the situation is really alarming (Sokolić, Rister, Luketin, 2009). Thus, it is high time to take steps towards change in the teaching methods and science promotion in the society.

OECD PISA (*Evolution of Student Interest in Science and Technology Studies*) results also indicate a need for an urgent reform in the Croatian educational system. PISA surveys are administered every three years on 15 year old high school students in 30 OECD member countries and 27 partner countries and economies, which together make up close to 90% of the world economy. Croatian students scored badly at the PISA testing:

– Croatia is at the bottom by the number of top-achievers in science and math;
– Croatia is at the very bottom by the number of students who have regular science classes at the school, as well as by the number of regular science classes held;
– Croatia is at the bottom by the science classes and science activities performed outside the school program;
– Croatian students named biology and astronomy as the most popular courses, but, out of 113.000 high school students nominally interested in astronomy only about 100 of them take part in the astronomy competition every year;
– PISA research also showed there is a misbalance in the development of schools in Croatia, which put Croatia at the bottom of the balanced European school systems (PISA research shows that socio-demographic background strongly affects the development of the schools and the students´ achievements);
– It is very worrisome that Croatian high school students already show a decrease in their interest in science career on the level comparable to the more developed countries, even though Croatia is economically significantly less developed.

As described earlier, Croatia needs more educated people, especially in the area of S&T. This can be achieved only if the youth become interested in science at an early age. Science fair competitions are not an ideal method of recruitment of young people into science and technology professions and identification of talented youth, but it is currently the only such an activity implemented through the whole Croatian school system and sponsored by the Croatian government (Vojnović, 2005.: 101). For this reason, science fair competitions for elementary and high school students are of an utmost importance for the development of Croatia. They recruit gifted students and provide a more challenging learning environment for top-achievers indicated in PISA. Unfortunately, Croatia does not have a system of identification and monitoring of gifted youth. Although some related documents[3] have been officially adopted, they have not been implemented yet. This is a tragic policy because, as emphasized by Vojnović (2005.: 101) , some research studies warn that children of high abilities, when incapacitated to develop their abilities, can manifest many shapes of the behavioural problems. For instance, in the USA there is a category of so called "black statistics" that indicate a good deal of the gifted people who ended up at the margins of the society. At the same time, USA has very well developed system for the gifted. Thus, if Croatia wants to break away from these negative trends, a real commitment is needed; youth can be motivated to take a part in S&T only under conditions that differ from the current educational system. We have to take into account that: I) children stabilize their area of interests around the age of 14; II) career choice is most frequently led by fantasies and this has to be taken seriously; III) children choose career in S&T because they consider it fascinating and they do not choose it because they consider it boring and IV) self-confidence is one of the crucial factors in the career choice (children often think they are not capable enough to work in S&T) (Schoon (2001); Osborne, Simon, Collins (2003)).

PISA gives a special emphasis to high-performers because "individuals with high level skills generate relatively large externalities in knowledge creation and utilisation, compared to an 'average' individual, which in turn suggests that investing in excellence may benefit all". Since Croatia has no programs for gifted or high-achieving students except science fair competitions, we decided to use these competitions as a proxy for exploring differences within the Croatian educational system. We explore the success of high school students from Croatian counties on the national level of science and mathematics competitions. A recent analysis of the astronomy competition by Vinković (2009.: 8) showed that the development index (DI) of Croatian counties relates to the success of the counties' students participating in the competition (Chart 1). But astronomy is a small competition, with only 197 high-school students entering the competition in 2009. Astronomy is also an elective course and not a part of the regular school curricula. This motivated us to try to get a broader perspective and make a survey on additional science competitions. Thus, we decided to include competitions in math, biology, chemistry and physics, which are often regular courses, and put them in relation to the counties' DI.

---

[3] *Pravilnik o osnovnoškolskom odgoju i obrazovanju darovitih učenika* (1991). Narodne novine, 34/1991. *Pravilnik o srednjoškolskom obrazovanju darovitih učenika* (1993). Narodne novine, 90/1993.

Two DIs were used in our analysis (Table 1):

1. "Regional Competitiveness index" (RCI) is produced following incentive of the National Competitiveness Council, the United Nations Development Programme and Croatian Chamber of Economy. Its elements are: A) business surrounding – demography, health and culture; education; basic infrastructure and public sector; business infrastructure; B) business sector – investments and entrepreneurial dynamics; development of the entrepreneurship; economic results – level; economic results – dynamics.
2. The second index is named "Socio-economic Index" (SEI) (Živić, Pokos, 2005). The indicators used in this index are: index of change of population from 1991 to 2001; index of ageing; coefficient of age dependence; rate of active population; employment rate; unemployment rate and "index of education". The indicators listed give us a basis for marking this index as index of more societal orientation than the previous, more business oriented index.

The success of a county in a science competition is measured by the number of students from a county who managed to reach the national level of competition (advancing first through the school level and then the county level). In order to avoid the size of the county affecting the result, we divided the number of students at the competitions (Table 2) with the expected number of students based on the total number of high-school students in the county (Table 1). This ratio between the actual and expected number of students is Science Fair Participation Index (SFPI). A balanced educational system would have SFPI close to one for all counties. In all the charts, the Y axis presents SFPI values, whereas the X axis presents one of the indexes – RCI or SEI. After forming the indexes tables we group the counties into three categories by development – low, medium and high developed counties, with 7 counties in each group – marked by error bars in the charts. Error bars represent one standard deviation σ of the group data:

$$\sigma = \sqrt{\sum_{i=1}^{N}(x_i - \langle x \rangle)^2 /(N-1)}$$

where $x_i$ are the data (RCI or SEI indexes or SFPI values) and $\langle x \rangle$ is their average (with N=7).

The available data on the science fair competitions were mostly from 2007-2009 and thus we decided to take this period as a referent one. But, we also had data on physics and mathematics from 2001-2009. These two disciplines suffer from a dramatic lack of teachers, which should affect the quality of physics and mathematics education. Indeed, as we will see, the older data for these disciplines are the ones that showed striking differences in the changes in SFPI within a decade, which should worry the stakeholders. For this reason, after presenting the data on biology and chemistry for the 2007-2009 period, we will show the data for physics and mathematics for 2007-2009, 2004-2006 and 2001-2003 periods.

Table 1.
Counties' rankings according to the Regional Competitiveness Index and Socio-economic Index and the number of high school students in the counties (a smaller Index means a higher economic development)

| County | Code | Regional Competitiveness Index | Socio-economic Index | Number of high school students | |
|---|---|---|---|---|---|
| | | | | total | within Croatia (%) |
| Bjelovar-Bilogora | BB | 11 | 9.4 | 5413 | 3.0 |
| Brod-Posavina | BP | 18 | 12.4 | 6981 | 3.9 |
| Dubrovnik-Neretva | DN | 10 | 10.3 | 5541 | 3.1 |
| City of Zagreb | GZ | 1 | 5.9 | 38594 | 21.3 |
| Istria | IS | 3 | 5.7 | 7956 | 4.4 |
| Karlovac | KA | 12 | 14.6 | 5112 | 2.8 |
| Koprivnica-Križevci | KK | 7 | 7.7 | 4645 | 2.6 |
| Krapina-Zagorje | KZ | 15 | 7.1 | 5343 | 2.9 |
| Lika-Senj | LS | 19 | 19.1 | 1479 | 0.8 |
| Međimurje | ME | 2 | 2.9 | 4282 | 2.4 |
| Osijek-Baranja | OB | 14 | 11.6 | 13820 | 7.6 |
| Požega-Slavonia | PS | 20 | 13.7 | 3882 | 2.1 |
| Primorje-Gorski Kotar | PG | 6 | 7.9 | 11387 | 6.3 |
| Šibenik-Knin | ŠK | 13 | 20.1 | 4669 | 2.6 |
| Sisak-Moslavina | SM | 16 | 16.7 | 5778 | 3.2 |
| Split-Dalmatia | SD | 8 | 10.1 | 21479 | 11.8 |
| Varaždin | VŽ | 4 | 7.0 | 8024 | 4.4 |
| Virovitica-Podravina | VP | 17 | 12.6 | 3738 | 2.1 |
| Vukovar-Syrmia | VS | 21 | 14.6 | 8023 | 4.4 |
| Zadar | ZD | 9 | 15.6 | 7659 | 4.2 |
| Zagreb County | ZG | 5 | 5.0 | 7505 | 4.1 |

Table 2.
The number of the high school students per county per scientific competition on the national level

| County | Astron. 2002-09 | Biology 2007-09 | Chemistry 2007-09 | Physics 2007-09 | Physics 2004-06 | Physics 2001-03 | Math. 2007-09 | Math. 2004-06 | Math. 2001-03 |
|---|---|---|---|---|---|---|---|---|---|
| BB | 2 | 15 | 9 | 1 | 0 | 0 | 0 | 0 | 2 |
| BP | 21 | 17 | 5 | 5 | 12 | 3 | 0 | 0 | 3 |
| DN | 3 | 3 | 5 | 12 | 11 | 10 | 8 | 8 | 6 |
| GZ | 99 | 97 | 86 | 97 | 72 | 81 | 131 | 101 | 112 |
| IS | 15 | 18 | 10 | 6 | 13 | 12 | 0 | 0 | 6 |
| KA | 5 | 16 | 10 | 6 | 3 | 1 | 0 | 1 | 1 |
| KK | 11 | 11 | 6 | 2 | 7 | 3 | 5 | 8 | 3 |
| KZ | 3 | 20 | 3 | 2 | 3 | 4 | 2 | 1 | 3 |
| LS | 0 | 0 | 1 | 1 | 1 | 2 | 0 | 1 | 0 |
| ME | 17 | 22 | 10 | 14 | 7 | 1 | 5 | 12 | 9 |
| OB | 19 | 17 | 10 | 16 | 5 | 9 | 12 | 15 | 11 |
| PS | 0 | 2 | 0 | 2 | 0 | 0 | 2 | 1 | 1 |
| PG | 10 | 8 | 13 | 10 | 12 | 10 | 21 | 25 | 37 |
| ŠK | 16 | 4 | 0 | 2 | 5 | 10 | 2 | 13 | 10 |
| SM | 27 | 20 | 5 | 4 | 6 | 6 | 0 | 4 | 7 |
| SD | 8 | 19 | 14 | 40 | 43 | 25 | 30 | 52 | 63 |
| VŽ | 6 | 25 | 17 | 18 | 12 | 12 | 18 | 16 | 2 |
| VP | 0 | 13 | 2 | 0 | 4 | 4 | 3 | 3 | 1 |
| VS | 0 | 1 | 16 | 2 | 3 | 1 | 2 | 5 | 4 |
| ZD | 1 | 14 | 10 | 6 | 12 | 13 | 5 | 21 | 22 |
| ZG | 10 | 13 | 1 | 0 | 2 | 4 | 3 | 6 | 4 |

Chart 1 shows the relation between the SFPI in astronomy and RCI, giving a clear, although slightly less pronounced trend than in the case of courses that will be discussed below. From the chart we can see that the SFPI follows a trend in RCI among the counties – apart from some exceptions like Šibenik-Knin County among medium developed and Sisak-Moslavina and Brod-Posavina among less developed. These counties are exceptionally high represented at the competitions. The counties that scored well in the astronomy competition also have good score at the RCI – especially Međimurje County, followed by Zagreb, Koprivnica-Križevci and Istria. Zagreb County rates well but Primorje-Gorski Kotar and Varaždin County count for well developed regions with low SFPI. Medium developed counties that scored poorly at the SFPI are Dubrovnik-Neretva, Bjelovar-Bilogora, Split-Dalmatia and Zadar County. The low developed counties whose scores are even worse (zero students at the competition) – Virovitica-Podravina, Lika-Senj, Požega-Slavonia and Vukovar-Syrmia – have to urgently change their practice in attracting students to the astronomy competition. Since astronomy is a small competition, the role of science fair mentors is highly pronounced. Less developed regions have a problem in attracting high-quality teachers who would prepare students for science competitions. This might be the trend that affects all competitions, as we will see below.

We showed astronomy as an example, without going into further clarifications. Astronomy is not a standard course at schools and in our paper we wanted to focus on regular courses and on the need of a change in the official educational system. Though, astronomy is not less important, it is meritorious for students' interest in the natural sciences (Sjøberg, Schreiner 2006; Vinković, 2009) and it should attract attention of education experts as well. Following the example of astronomy competition, we now present the results of comparing the SFPI with both the RCI and the SEI for other competitions.

Chart 1.
*Relation of astronomy competition results (averaged over the years from 2002 to 2009) and Regional Competitiveness Index*

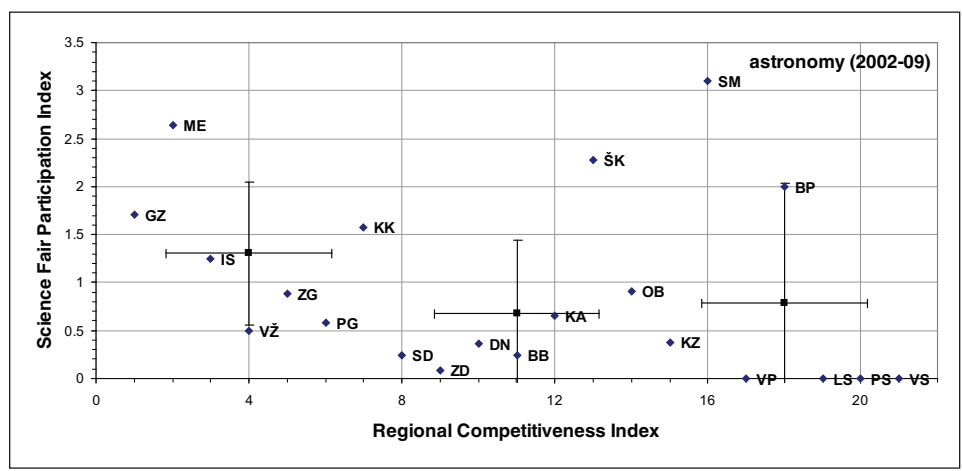

## 2. a) Biology

Biology science fair competition was the second largest among our five – 2939 students took part in 2009. We elaborated that our two development indexes comprise different indicators and thus the differences in their relation to the SFPI can appear. This is the case in biology – SEI better indicates a trend than RCI. Speaking about single indexes, RCI (Chart 2) tells us that around half of the counties follow their RCI when it comes to the SFPI – Međimurje, Varaždin and Zagreb and to some extent Istria and Koprivnica-Križevci (among the high developed). Counties that score exceptionally well when compared to their low RCI are: Krapina-Zagorje, Sisak-Moslavina, Virovitica-Podravina and Brod-Posavina. Medium ranked counties with very good results at SFPI are Karlovac and Bjelovar-Bilogora, while counties like Split-Dalmatia, Dubrovnik-Neretva, Šibenik-Knin and Osijek-Baranja present the medium developed counties with very low score on the SFPI.

Very similar trends are present for the SEI (Chart 3), with a difference that some medium and low developed counties have inversed positions on these two indexes. In both indexes Međimurje and Krapina-Zagorje stand out as positive examples of high or medium developed counties with a decent SFPI while Vukovar-Syrmia and Lika-Senj are definitely the counties that have to invest in attracting science fair mentors and change their absence from the biology competition.

Chart 2.
*Relation of biology competition results and Regional Competitiveness Index (2007-2009)*

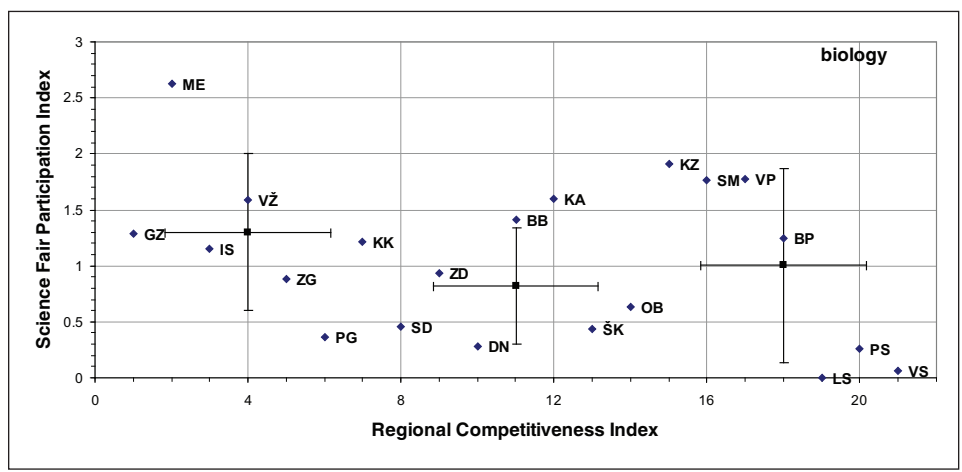

Chart 3.
*Relation of biology competition results and Socio-economic Index (2007-2009)*

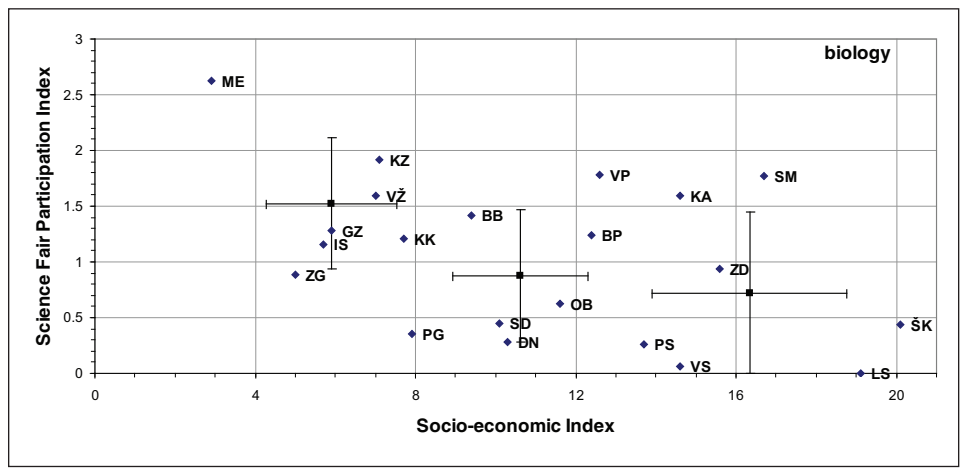

## 2. b) Chemistry

In 2009 chemistry competition was the third by size among the observed competitions – 2024 students participated. Comparing the two development indexes, we can say that in the case of chemistry the RCI (Chart 4) was more "strict" one, placing more counties with low SFPI into the low developed group than the SEI (Chart 5) does it. Equally to the results from biology competition, Međimurje, Varaždin and Zagreb are again highly developed areas with very high score on SFPI. On the other hand, Zagreb County stands out as a highly developed region but with a very small SFPI. Split-Dalmatia is a medium developed county with low SFPI, Dubrovnik-Neretva is trying to catch on while Šibenik-Knin is an extreme case with no students at the competition. Požega-Slavonia among low developed has no students, whereas Sisak-Moslavina, Osijek-Baranja, Brod-Posavina, Lika Senj, Virovitica-Podravina and Krapina-Zagorje areas are low ranked in RCI but has at least some representation at the competitions. Karlovac and Bjelovar-Bilogora are among medium developed counties with high SFPI. Vukovar-Syrmia is a surprise here as a low developed county with very high SFPI.

In the case of chemistry, SEI follows the logic of RCI, except that some counties are not among low developed, but among medium or high developed, and this changes the picture slightly. Osijek-Baranja, Brod-Posavina, Virovitica-Podravina are now among the medium developed and Krapina-Zagorje is the county placed now among the high developed county.

Chart 4.
Relation of chemistry competition results and Regional Competitiveness Index (2007-2009)

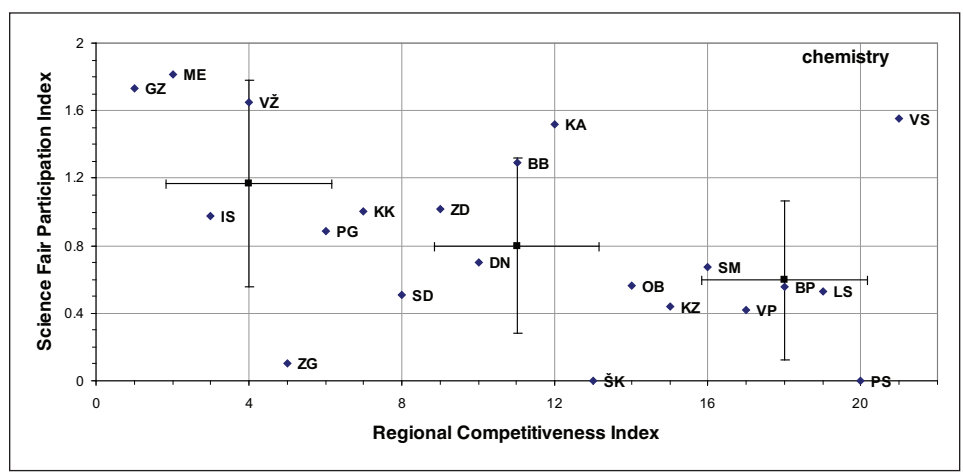

Chart 5.
*Relation of chemistry competition results and Socio-economic index (2007-2009)*

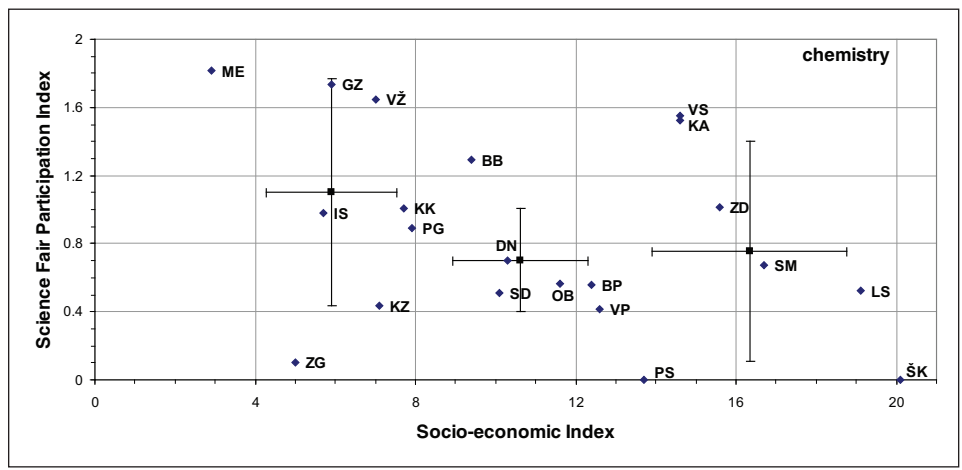

## 2 c) Physics

After astronomy physics was the smallest science fair competition in 2009 – 1588 students took part in it. As announced earlier on, data on competition in physics were available for 2001-2009 period and here we bring the charts presenting the data in chronological order. We will not interpret the data for 2001-2003 and 2004-2006 periods in detail. These earlier periods will serve us only as a proof of rapid change in a direction of stronger inter-correlation of the participation of students in the science fair competitions and the development index. In other words, these earlier periods (Charts 6-9) show us that the differences among the counties were not so striking as in 2007-2009 period; social differentiation and differences in quality of science teaching have taken place in a very recent period and it is a trend that has a chance of being stopped if addressed immediately.

**Chart 6.**
Relation of physics competition results and Regional Competitiveness Index (2001-2003)

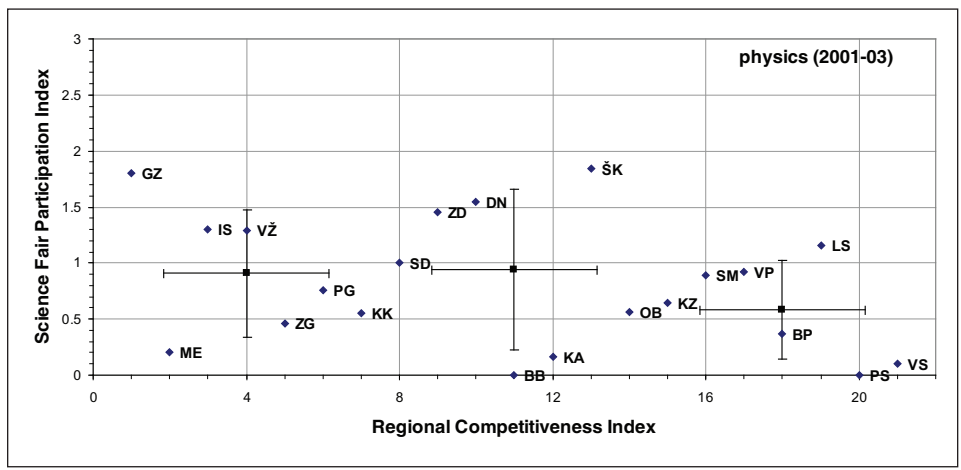

**Chart 7.**
*Relation of physics competition results and Socio-economic index (2001-2003)*

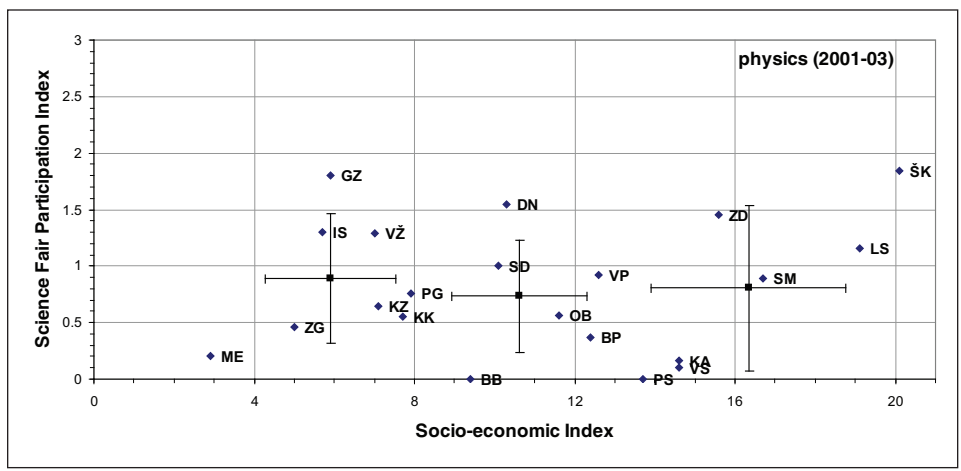

Chart 8.
*Relation of physics competition results and Regional Competitiveness Index (2004-2006)*

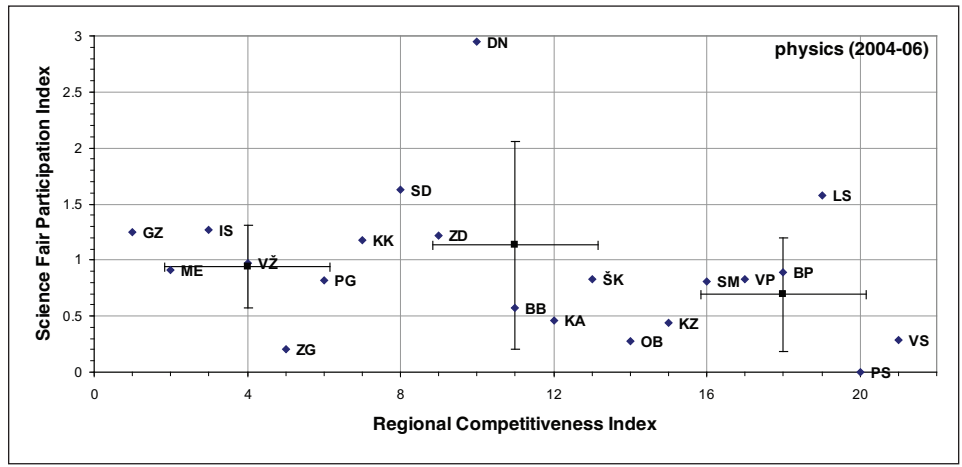

Chart 9.
*Relation of physics competition results and Socio-economic index (2004-2006)*

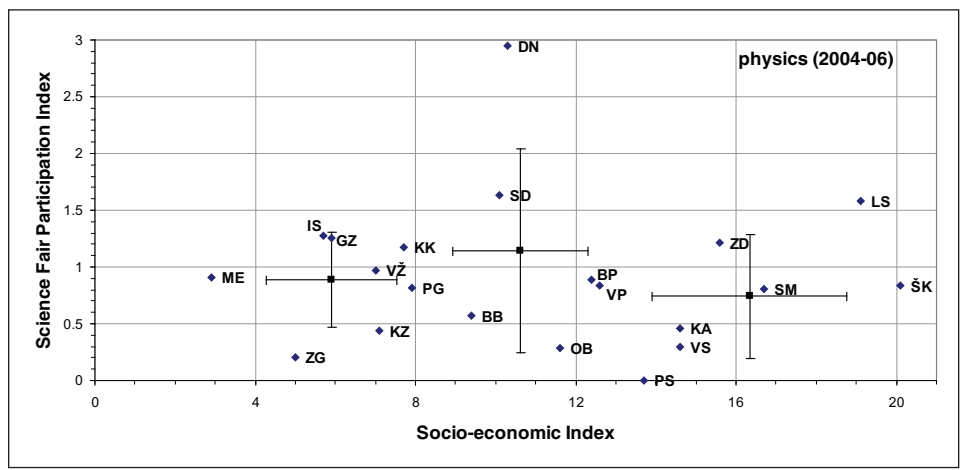

Starting from Chart 10 (RCI) and Chart 11 (SEI), a clear trend is visible in relation of the students' participation in physics competition and the development index. In the case of counties that outnumber in SFPI and have a good RCI/SEI, physics follows examples of biology and chemistry; Međimurje, Zagreb and Varaždin have the best results, whereas their counterparts in development, but with a low score on SFPI are Istria, Primorje-Gorski Kotar, Koprivnica-Križevci, (and Krapina Zagorje in the case of SEI) and Zagreb County with no students. Medium developed counties highly present at the physics competition are Dubrovnik-Neretva and Split-Dalmatia, while Zadar, Šibenik-Knin and Bjelovar-Bilogora are the medium developed counties according to RCI, but with a low SFPI. Counties that are low developed and have poor presentation at the physics competition are Brod-Posavina, Sisak-Moslavina, Lika-Senj and Požega-Slavonia whereas Vukovar-Syrmia is barely present and Virovitica-Podravina is not present at all.

It is interesting to note that Međimurje had a remarkable rise from one of the least successful counties in 2001-2003 to the best county in 2007-2009. On the other hand, Šibenik-Knin County was the best in 2001-2003 but dropped into the group with a low SFPI in 2007-2009. We assume that these trends are the result of a group of mentors who appear or disappear over the period of one decade. When these mentors stop their activities the whole system of science fair support in a county collapses. This is a sobering warning that a network of well trained and committed mentors is needed instead of individuals who act alone.

Chart 10.
*Relation of physics competition results and Regional Competitiveness Index (2007-2009)*

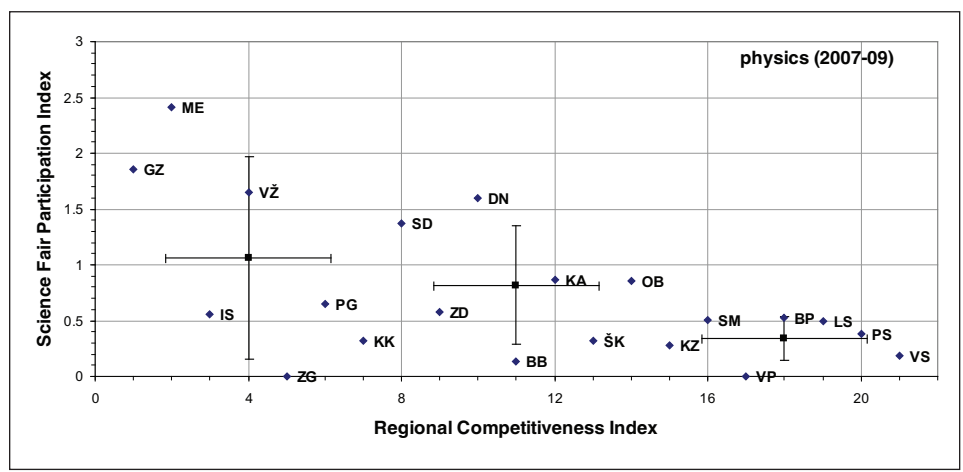

Chart 11.
*Relation of physics competition results and Socio-economic index (2007-2009)*

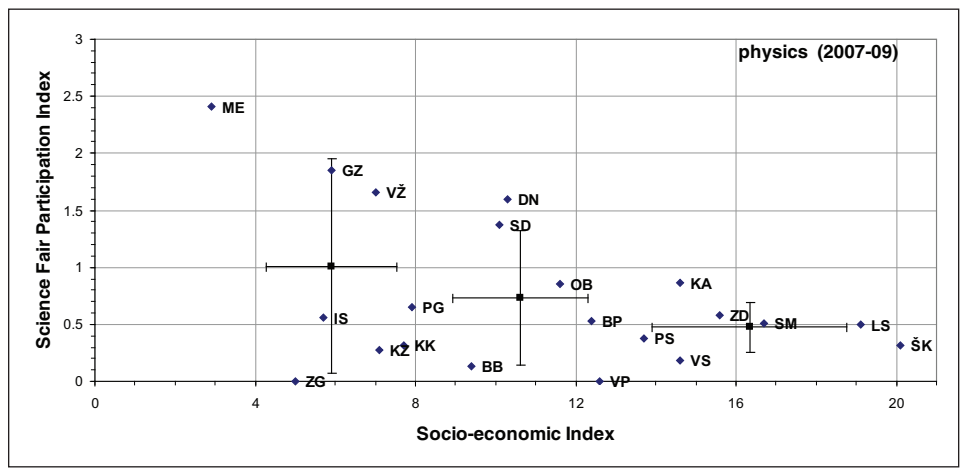

## 2. d) Mathematics

Mathematics is the largest competition observed here[4]; it had 3703 participants in 2009. Its data is also available from 2001-2009 and earlier periods will serve again only as an indication that the changes in social differentiation and greater differences in a rate of students taking part in the math science fair competitions occurred only recently. Evidence for this statement can be deduced from the data presented in the charts from 12-15; a slight trend can be observed only for 2004-2006 period, but it is only in 2007-2009 period that the trend becomes more pronounced and offers a basis for more sound conclusions.

---

[4] Mathematics competitions changed their rules in 2006 when two competition categories were introduced: A for high schools with a special natural science curriculum (prirodoslovno-matematičke gimnazije) and B for all other schools. Students from these other schools were allowed to compete in category A, but not the vice versa. Since the invitation to the national competition is offered to all the winners of the county levels in category B, we used data for category A where only the top 20 students from the county level advance to the national level.

**Chart 12.**
*Relation of mathematics competition results and Regional Competitiveness Index (2001-2003)*

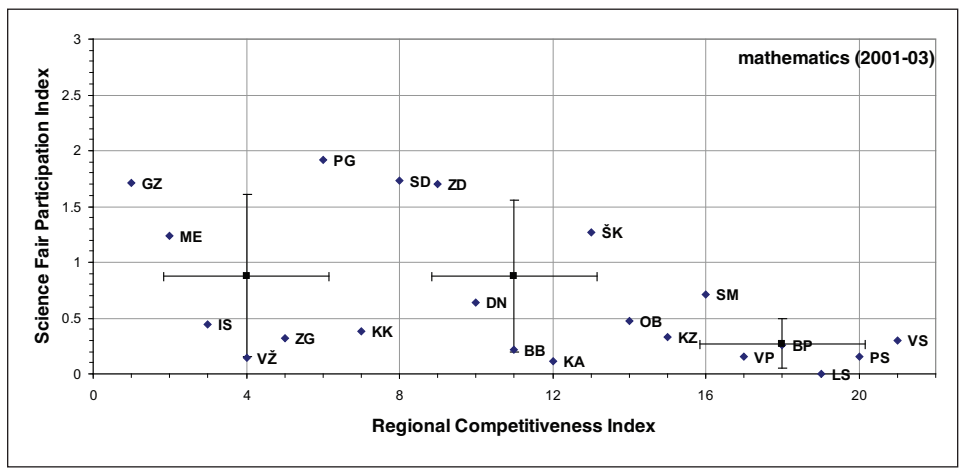

**Chart 13.**
*Relation of mathematics competition results and Socio-economic index (2001-2003)*

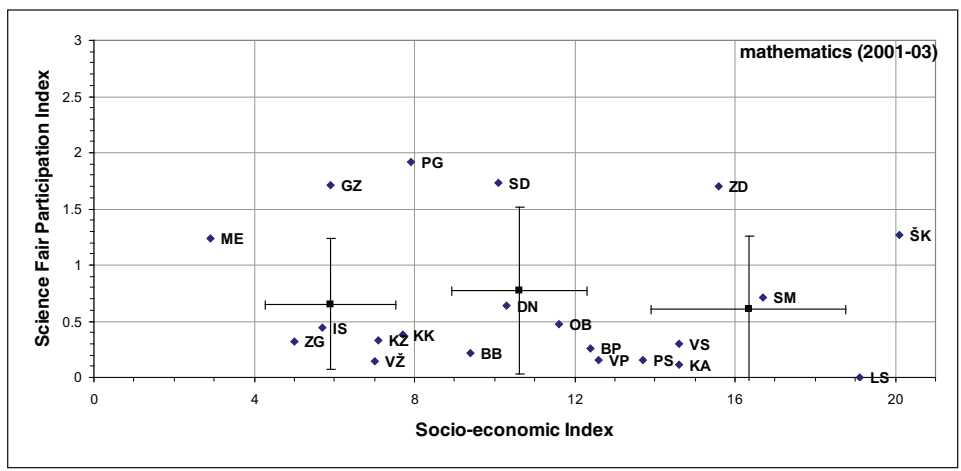

*D. Vinković, D. Potočnik: Educational asymmetries in the making: Science Fair Competitions...*

**Chart 14.**
*Relation of mathematics competition results and Regional Competitiveness Index (2004-2006)*

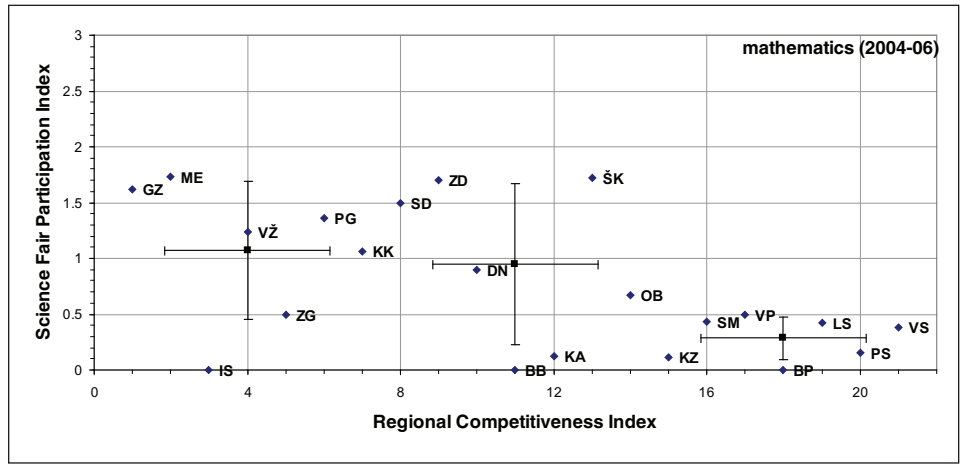

**Chart 15.**
*Relation of mathematics competition results and Socio-economic index (2004-2006)*

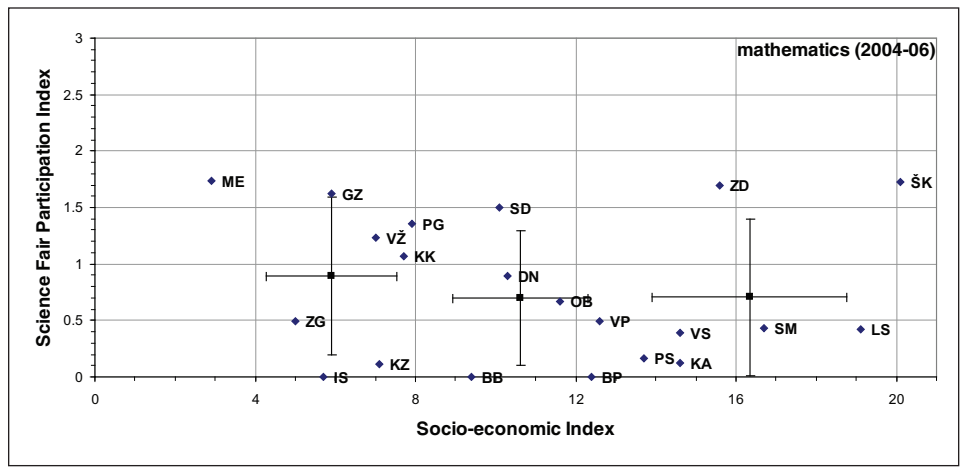

Sociologija i prostor

69

Although math is a competition with the highest number of participants among natural sciences, there are counties with no student in the competition on the national level – this is the case for Istria, Bjelovar-Bilogora, Karlovac, Sisak-Moslavina, Brod-Posavina and Lika-Senj. The rest of the counties in both indexes – RCI (Chart 16) and SEI (Chart 17) – in the case of well developed counties show that Zagreb is outscoring all counties, followed by Varaždin and Primorje-Gorski Kotar. On the other side, both indexes show there are well developed counties with low SFPI, like Zagreb County and Istria. In the case of RCI, Zadar, Šibenik-Knin and Osijek-Baranja are the counties with medium development index and low representation at the math competition, while the same is with Osijek-Baranja and Virovitica-Podravina in the case of SEI. The results in RCI comparison show Krapina-Zagorje, Požega-Slavonija and Vukovar-Syrmia as low developed counties with a low SFPI score. The same is the case with Požega-Slavonija, Zadar, Šibenik-Knin and Vukovar-Syrmia when their SFPI is compared with the SE index.

Chart 16.
*Relation of mathematics competition results and Regional Competitiveness Index (2007-2009)*

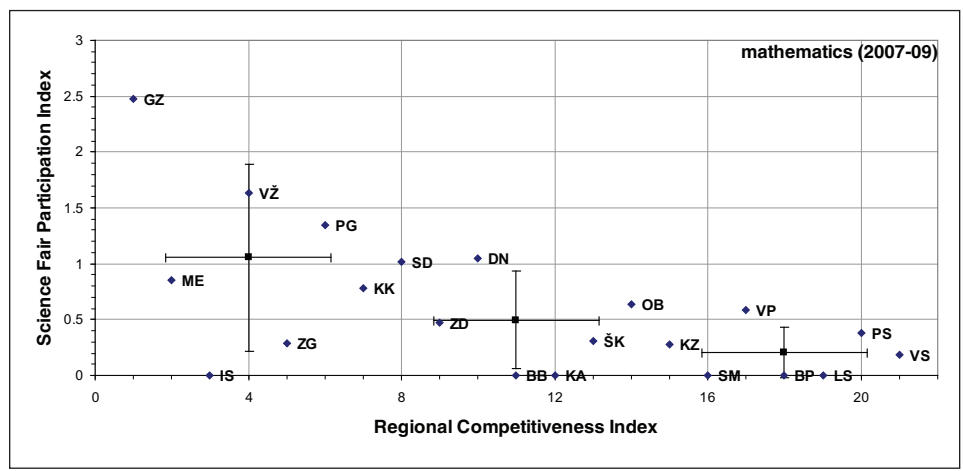

Chart 17.
*Relation of mathematics competition results and Socio-economic index (2007-2009)*

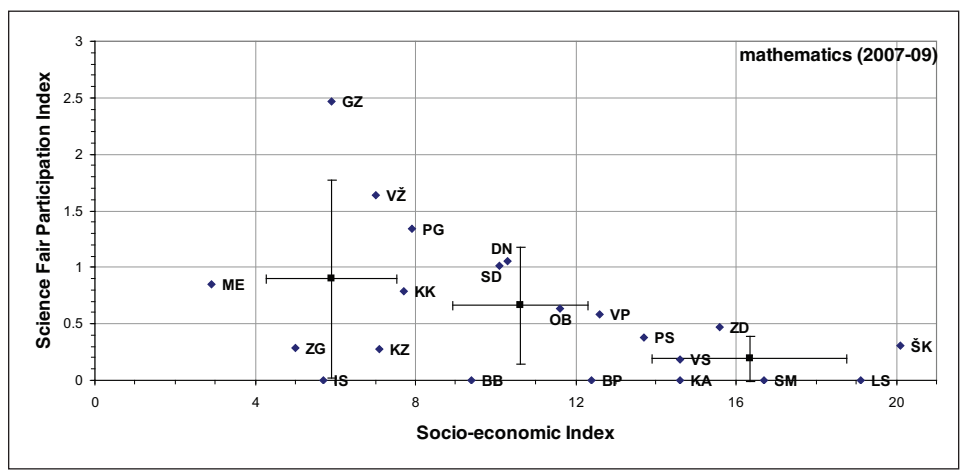

## 3. Conclusions and recommendations

Our research was motivated by the results from the astronomy science fair competition report (Vinković, 2009) whose conclusion is that the counties' development index is related to the participation rate of the students at the competitions. This finding has broader implications than those of the popularity of science fair competition system; low students' interest in science implies low enrolment in natural science study and low employment in R&D sector, which results in prolongation of underdevelopment of these regions. Thus, we expanded this survey and aimed at comparing the Science Fair Participation Index (SFPI) in biology, chemistry, physics and mathematics with two development indexes. One is "Regional Competitiveness index" comprising broader business environment and business sector itself and another is "Socio-economic Index" which consists of population and employment indicators. Referent period for comparison was 2007-2009 for all disciplines, adding 2001-2003 and 2004-2006 for physics and mathematics. When compared with these two indexes, SFPI gives the following results (SFPI equal one means the number of science competition participants is as expected from the county size):

1. The City of Zagreb, Varaždin and Međimurje have the highest score in the development indexes and the highest SFPI in almost all of the five observed disciplines;

2. Međimurje County is a unique example as its SFPI is very high at astronomy, biology, chemistry and physics and with SFPI of about one in mathematics;

3. Virovitica-Podravina, Lika-Senj, Požega-Slavonija and Vukovar-Syrmia are in almost all examples the counties with the low development indexes and low SFPI – these counties have to change their practice in attracting mentors and students to the science competitions urgently;

4. There are counties with no students at the competitions during 2007-2009 and they are:

   – for astronomy: Virovitica-Podravina, Lika-Senj, Požega-Slavonia and Vukovar-Syrmia for astronomy, Lika-Senj for biology, Šibenik-Knin and Požega-Slavonia for chemistry, Bjelovar-Bilogora and Požega-Slavonia for physics and Istria, Bjelovar-Bilogora, Karlovac, Sisak-Moslavina, Brod-Posavina and Lika-Senj for mathematics.

Following the results of our analysis, we can stress out three broad recommendations for improvement of science fair competition system and fostering pupils' and students' interest in science and technology:

1. Clear goals, aims and long-term strategy of the science fair competition system, in accordance to the existing educational and economic strategies have to be defined.

2. Sociological researches of interest, motivations and socio-demographic factors of the students and their mentors have to be conducted on a regular basis.

3. Cooperation between universities, scientific institutions and scientists on one side and science fair competitions on the other has to be established as a regular practice. Compulsory scientific outreach should be introduced in project planning of the publicly financed S&T research projects.

This paper aims to provide basis for profound conclusions on urgent needs for a reform of science education, science competitions, and cooperation of the science and society. Much can be done on a county level, too, where a support can be given to science fair mentors and outreach projects. It remains to be seen whether the stakeholders in political, economic and educational sphere will find it in their power to act upon these messages on the development and prospects of Croatia.

### X. Acknowledgments

We thank Duje Bonacci, Antonela Nižetić-Capković, Neda Lesar, Ela Rac-Marinić-Kragić and Antonija Horvatek for helping us collect the data for this study.

### Literature

1. Babić, Z.; Matković, T.; Šošić, V. (2006). Structural Changes in Tertiary Education and Impacts on the Labor Market. *Privredna kretanja i ekonomska politika* 108: 26-66.
2. Baranović, B. (Ed.) (2006). *Nacionalni kurikulum za obvezno obrazovanje u Hrvatskoj: različite perspektive*. Zagreb: Institut za društvena istraživanja.
3. Bejaković, P.; Lowther, J. (2005). *The competitiveness of Croatia's human resources*. Zagreb: Institute of Public Finance.

4. Ederer, P.; Schuler, P.; Willms, S. (2007). *Lisbon Council Policy Brief: European Human Capital Index, Central and Eastern Europe* http://www.zeppelinuniversity.de/deutsch/scientific_services/TheEuropeanHumanCapitalIndex.pdf
5. *Europe in Figures* (2008). Eurostat Yearbook http://epp.eurostat.ec.europa.eu/cache/ITY_OFFPUB/KS-CD-07-001-12/EN/KS-CD-07-001-12-EN.PDF
6. *Europe needs more scientists* (2004). Report by the High Level Group on Increasing Human Resources for Science and Technology in Europe. European Commission http://ec.europa.eu/research/press/2004/pr0204en.cfm
7. *European Research in Action: Young People and Science* http://ec.europa.eu/research/leaflets/young/index_en.html
8. *Europeans, Science & Technology* (2005). Special EUROBAROMETER 224. European Commission http://ec.europa.eu/public_opinion/archives/ebs/ebs_224_report_en.pdf
9. *Evolution of Student Interest in Science and Technology Studies* (2006). European Science Forum. OECD http://www.oecd.org/dataoecd/16/30/36645825.pdf
10. *Koncept: Odkrivanje in delo z nadarjenim učenci* (1999). Nacionalni kurikularni svet, Področna kurikularna komisija za osnovno šolo, Delovna skupina za pripravo koncepta dela z nadarjenimi učenci. Ministarstvo za šolstvo in šport Republike Slovenije http://www.mss.edus.si
11. Lent, R. W.; Brown, S. D.; Hackett, G. (2002). Social cognitive career theory. In: Brown, D. et al. (Eds.). *Career choice and development.* San Francisco: Jossey-Bass.: 255-311.
12. Lučin, P. (2007). *Strateški okvir razvoja znanosti do 2010*. Drugi kongres hrvatskih znanstvenika iz domovine i inozemstva.
13. Meri T. (2008). *Statistics in Focus* #18. Eurostat www.estatisticas.gpeari.mctes.pt/archive/doc/High-tech_knowledge_intensive_services_0.PDF
14. Osborne, J.; Simon, S.; Collins, S. (2003). Attitudes towards science: a review of the literature and its implications. *International Journal of Science Education*, (25): 9:1049-1079.
15. *PISA 2006: Science Competencies for Tomorrow's World* (2006). Volume 1: Analysis http://www.pisa.oecd.org/dataoecd/30/17/39703267.pdf
16. *Popis stanovništva* (2001). Državni zavod za statistiku http://www.dzs.hr/Hrv/censuses/Census2001/Popis/H01_01_07/H01_01_07.html
17. Potočnik, D. (2009). Izbor studija: motivacijska struktura upisa i očekivani uspjeh u pronalasku posla. *Sociologija i prostor* 46(3-4): 265-284.
18. *Pravilnik o osnovnoškolskom odgoju i obrazovanju darovitih učenika* (1991). Narodne novine, 34/1991.
19. *Pravilnik o srednjoškolskom obrazovanju darovitih učenika* (1993). Narodne novine, 90/1993.
20. *Consultation on the Future "EU 2020" Strategy* (2009). European Commission http://ec.europa.eu/eu2020/pdf/eu2020_en.pdf
21. *Science and Technology Policy of the Republic of Croatia 2006-2010* (2006) Republic of Croatia, Ministry of Science, Education and Sports http://public.mzos.hr/fgs.axd?id=14189
22. *Science Education NOW: A renewed Pedagogy for the Future of Europe* (2007). European Commission http://ec.europa.eu/research/science-society/document_library/pdf_06/report-rocard-on-science-education_en.pdf

23. Schoon, I. (2001). Teenage job aspirations and career attainment in adulthood: A 17-year follow-up study of teenagers who aspired to become scientists, health professionals, or engineers. *International Journal of Behavioral Development*, 25(2): 124-132.
24. Sjøberg, S.; Schreiner, C. (2006). How do learners in different cultures relate to science and technology? Results and perspectives from the project ROSE (the Relevance of Science Education). *APFSLT: Asia-Pacific Forum on Science Learning and Teaching*, 7(1), Foreword.
25. Sokolić, F.; Rister, D.; Luketin, I. (2008). *The Role of the History of Science in Physics Education, Teachers' Vision* http://bib.irb.hr/prikazi-rad?&rad=396299
26. *Strateški okvir za razvoj 2006.-2013.* (2006). Vlada Republike Hrvatske. http://www.vlada.hr/hr/preuzimanja/publikacije/strateski_okvir_za_razvoj_2006_2013
27. *The Global Competitiveness Report* (2009). World Economic Forum http://www.weforum.org/en/initiatives/gcp/Global%20Competitiveness%20Report/index.htm
28. *The Lisbon Strategy* (2000). European Council http://ec.europa.eu/archives/ISPO/docs/services/docs/2000/jan-march/doc_00_8_en.html#A
29. *Top of the Class - High Performers in Science in PISA 2006* (2009). OECD http://www.pisa.oecd.org/document/51/0,3343,en_32252351_32236191_42642227_1_1_1_1,00.html
30. Vinković, D. (2009). *Izvještaj predsjednika Državnog povjerenstva za provedbu i organizaciju natjecanja iz astronomije 2008/2009.: Natjecanja iz astronomije: analiza stanja i prijedlozi za daljnji razvoj* http://vinkovic.org/Projects/MindExercises/astronomija/Izvjestaj_natjecanje_astronomija.pdf
31. Vlahović – Štetić, V. (Ed.) (2005). *Daroviti učenici: teorijski pristup i primjena u školi*. Zagreb: Institut za društvena istraživanja.
32. Vojnović, N. (2005). Stanje, problemi i potrebe u području skrbi o darovitim učenicima u Hrvatskom školskom sustavu. In: Vlahović – Štetić, V. (Ed.) (2005). *Daroviti učenici: teorijski pristup i primjena u školi*. Zagreb: Institut za društvena istraživanja, 71-107.
33. *World Economic Forum. Rankings: Global Competitiveness Report 2008-2009* http://www.weforum.org/documents/gcr0809/index.html
34. *Znanstvena i tehnologijska politika Republike Hrvatske 2006.-2010.* (2006). Vlada Republike Hrvatske http://www.vlada.hr/hr/preuzimanja/publikacije/znanstvena_i_tehnologijska_politika_republike_hrvatske_2006
35. Živić, D.; Pokos, N. (2005). Odabrani sociodemografski indikatori razvijenosti Hrvatske i županija. *Revija za sociologiju*, 36(3-4): 207-224.

*D e j a n   V i n k o v i ć*
*Prirodoslovno Matematički Fakultet, Sveučilište u Splitu*
vinkovic@pmfst.hr

*D u n j a   P o t o č n i k*
*Institut za Društvena istraživanja u Zagrebu*
dunja@idi.hr

**Obrazovne asimetrije u nastajanju: natjecanja učenika kao indikatori ekonomskog razvoja**

**Sažetak**

Hrvatska ne stoji dobro na ljestvici indeksa razvijenosti, stope zaposlenosti, razvoja visokotehnološkog sektora, što su samo neki od indikatora koji ne daju perspektivu promjene na bolje. Također, hrvatski srednjoškolci su na dnu ljestvice u Europi po rezultatima na testovima znanosti, a opada i njihov interes za karijeru u znanosti. Hrvatska treba više obrazovanih ljudi, posebice u području znanosti i tehnologije, što se može postići jedino ukoliko se mlade u najranijoj dobi zainteresira za znanost. Stoga su natjecanja u pojedinim znanstvenim disciplinama od presudne važnosti za razvoj Hrvatske. Ovaj članak cilja ka istraživanju povezanosti indeksa razvijenosti županija sa stopom sudjelovanja učenika na natjecanjima. Navedeno će se postići uključenjem dvaju indeksa razvijenosti u traženje povezanosti sa stopama sudjelovanja u natjecanjima iz biologije, kemije, fizike i matematike, uz pregled prethodnog istraživanja za natjecanje iz astronomije. Kao što će se pokazati, postoji rastući trend međusobne povezanosti indeksa razvijenosti i stopa sudjelovanja učenika na natjecanjima po svim disciplinama.

*Ključne riječi:*  uspjeh učenika na natjecanjima, socio-ekonomski indeks razvijenosti, motivacija učenika, interes učenika za prirodne znanosti.